\documentclass[preprint,10pt,twocolumn,tightenlines,showpacs,amsmath,amssymb,twoside]{revtex4-1}
\usepackage{epsfig}
\usepackage{dcolumn}
\usepackage{bm}
\usepackage{color}
\usepackage{graphicx}
\usepackage{subfigure}
\usepackage{pstricks}
\usepackage{pst-node}
\usepackage{rotating}
\usepackage{times}
\usepackage[english]{babel}
\addto{\captionsenglish}{%

}
\usepackage[colorlinks,linkcolor=blue,citecolor=red]{hyperref}
\usepackage{lineno}

\newcommand{\chicJ}{\chi_{cJ}}

\newcommand{\ssb}{\Sigma^0\bar{\Sigma}^0}
\newcommand{\ccb}{c\bar{c}}

\newcommand{\XXB}{\Xi^{-}\bar\Xi^{+}}

\newcommand{\SSSM}{\Sigma(1385)^{-}\bar\Sigma(1385)^{+}}
\newcommand{\SSSP}{\Sigma(1385)^{+}\bar\Sigma(1385)^{-}}
\newcommand{\SSPM}{\Sigma(1385)^{\mp}\bar\Sigma(1385)^{\pm}}
\newcommand{\EE}{e^+e^-}

\newcommand{\BB}{B\bar{B}}
\newcommand{\ppb}{p\bar{p}}

\newcommand{\psp}{\psi(3686)}
\newcommand{\jpsi}{J/\psi}
\newcommand{\ar}{\rightarrow}

\newcommand{\llb}{\Lambda\bar{\Lambda}}

\newcommand{\bfg}{\begin{figure}}
\newcommand{\efg}{\end{figure}}
\newcommand{\bitm}{\begin{itemize}}
\newcommand{\eitm}{\end{itemize}}
\newcommand{\bnum}{\begin{enumerate}}
\newcommand{\enum}{\end{enumerate}}
\newcommand{\btbl}{\begin{table*}}
\newcommand{\etbl}{\end{table*}}
\newcommand{\btbu}{\begin{tabular}}
\newcommand{\etbu}{\end{tabular}}
\newcommand{\bcl}{\begin{center}}
\newcommand{\ecl}{\end{center}}
\newcommand{\bbt}{\bibitem}
\newcommand{\beq}{\begin{equation}}
\newcommand{\eeq}{\end{equation}}
\newcommand{\beqr}{\begin{eqnarray}}
\newcommand{\eeqr}{\end{eqnarray}}
\begin{document}
\normalsize
\parskip=5pt plus 1pt minus 1pt
\title{\boldmath Study of $\psi$ decays to the $\XXB$ and $\SSPM$ final states}
\author{ 
  M.~Ablikim$^{1}$, M.~N.~Achasov$^{9,e}$, X.~C.~Ai$^{1}$,
  O.~Albayrak$^{5}$, M.~Albrecht$^{4}$, D.~J.~Ambrose$^{44}$,
  A.~Amoroso$^{49A,49C}$, F.~F.~An$^{1}$, Q.~An$^{46,a}$,
  J.~Z.~Bai$^{1}$
% ...
  \\
  \vspace{0.2cm}
  (BESIII Collaboration)\\
  \vspace{0.2cm} {\it
    $^{1}$ Institute of High Energy Physics, Beijing 100049, People's Republic of China\\
% ...
$^{4}$ Bochum Ruhr-University, D-44780 Bochum, Germany\\
    $^{5}$ Carnegie Mellon University, Pittsburgh, Pennsylvania 15213, USA\\
% ...
$^{9}$ G.I. Budker Institute of Nuclear Physics SB RAS (BINP), Novosibirsk 630090, Russia\\
% ...
$^{44}$ University of Rochester, Rochester, New York 14627, USA\\
% ...
$^{46}$ University of Science and Technology of China, Hefei 230026, People's Republic of China\\
% ...
$^{49}$ (A)University of Turin, I-10125, Turin, Italy; (B)University of Eastern Piedmont, I-15121, Alessandria, Italy; (C)INFN, I-10125, Turin, Italy\\
% ...
$^{a}$ Also at State Key Laboratory of Particle Detection and Electronics, Beijing 100049, Hefei 230026, People's Republic of China\\
% ...
$^{e}$ Also at the Novosibirsk State University, Novosibirsk, 630090, Russia\\
% ...
  }
}

\vspace{1.4cm}
%\date{\today}
\begin{abstract}
We study the decays of the charmonium resonances $\jpsi$ and $\psp$ to the final states $\XXB$, $\SSPM$ based on a single baryon tag method using data samples of $(223.7 \pm 1.4) \times 10^{6}$ $\jpsi$ and $(106.4 \pm 0.9) \times 10^{6}$ $\psp$ events collected with the BESIII detector at the BEPCII collider.
The decay $\psp\ar\SSPM$ is observed for the first time, and the measurements of the other processes, including the branching fractions and angular distributions, are in good agreement with, and much more precise than, the previously published results.
Additionally, the ratios $\frac{{\cal{B}}(\psp\ar\XXB)}{{\cal{B}}(\jpsi\ar\XXB)}$, $\frac{{\cal{B}}(\psp\ar\SSSM)}{{\cal{B}}(\jpsi\ar\SSSM)}$ and $\frac{{\cal{B}}(\psp\ar\SSSP)}{{\cal{B}}(\jpsi\ar\SSSP)}$ are determined.
\end{abstract}
\maketitle

%\begin{multicols}{2}
%\linenumbers
\section{Introduction}
The study of $\psi$ [in the following, $\psi$ denotes both charmonium resonances $\jpsi$ and $\psp$] production in $\EE$ annihilation and the subsequent two-body hadronic decays of the $\psi$, such as baryon-antibaryon decays, provide a unique opportunity to test quantum chromodynamics (QCD) in the perturbative energy regime and to study the baryonic properties~\cite{Farrar}.
These decays are expected to proceed via the annihilation of $\ccb$ into three gluons or a virtual photon. This model also leads to the prediction that the ratio of the branching fractions of $\psi$ decays to a
specific final state should follow the so-called \textquotedblleft 12\% rule\textquotedblright~\cite{12rule}
\begin{equation}
\frac{{\cal B}(\psp\ar \text{hadrons})}{{\cal B}(J/\psi\ar \text{hadrons})} \approx
 \frac{{\cal B}(\psp\ar\EE)}{{\cal B}(J/\psi\ar\EE)} \approx 12\%,
\end{equation}
where the branching fractions probe the ratio of the wave functions at their origins for the vector ground state $\jpsi$ and its first radial excitation $\psp$.
This rule was first observed to be violated in the process $\psi\ar\rho\pi$, which is known as the \textquotedblleft $\rho\pi$ puzzle,\textquotedblright
and was subsequently further tested in a wide variety of experimental measurements~\cite{PDG2012}.
Recently, a review of the theoretical and experimental results~\cite{yfgu} concluded
that the current theoretical explanations are unsatisfactory, especially for the baryon pair decays of $\psi$ mesons.
Therefore, more experimental measurements on baryon-antibaryon ($\BB$) pair final states,
e.g. $\ppb, \llb, \Sigma\bar\Sigma, \Xi\bar\Xi, \Sigma(1385)\bar\Sigma(1385)$, in the decays of $\psi$ are desirable.
To date, the branching fractions of the decays $\psi\ar\XXB$ and $\jpsi\ar\SSPM$ were previously measured with a low precision~\cite{Jximxip01,Jximxip02,Jximxip03,Jximxip04,Pximxip},
and the decay $\psp\ar\SSPM$ has not yet been observed.

By using hadron helicity conservation,
the angular distribution for the process $\EE\ar\psi\ar\BB$ can be expressed as
\begin{equation}\label{alpha}
\frac{dN}{d(\cos\theta)}\propto1+\alpha\cos^{2}\theta,
\end{equation}
where $\theta$ is the angle between the baryon and the positron-beam direction in the $\EE$ center-of-mass (CM) system and $\alpha$ is a constant.
Various theoretical calculations based on first-order QCD have made predictions for the value of $\alpha$.
In the prediction of Claudson {\it et al.}~\cite{ppbref02}, the baryon mass is taken into account
as a whole, while the constituent quarks inside the baryon are considered as massless when computing the decay amplitude.
The prediction by Carimalo~\cite{ppbref01} takes the mass effects at the quark level into account.
Experimental efforts are useful to measure $\alpha$ in order to test the hadron helicity
conservation rule and study the validity of the various theoretical approaches.
In the previous experiments, the angular distributions are measured with a few decays,
such as $\psp\ar\ppb$~\cite{ppb} and $\jpsi\ar\BB$ $[\ppb, \llb, \ssb, \XXB, \Sigma(1385)\bar\Sigma(1385)$]~\cite{Jximxip04,ppbref,angularSig,abc}. Among them, the angular distributions for the $\jpsi\ar\XXB, \SSPM$ decays are determined with a low precision, while for the decays $\psp\ar\XXB$, $\SSPM$ have not yet been measured.

In this paper, we report the most precise measurements of the branching fractions and angular distributions for the decays $\psi\ar\XXB$, $\SSPM$
based on $(223.7 \pm 1.4) \times 10^{6}$ $J/\psi$~\cite{Jpsi} and $(106.4 \pm 0.9) \times 10^{6}$ $\psp$~\cite{Psip} events collected
with the BESIII detector at BEPCII.

\section{BESIII DETECTOR AND MONTE CARLO SIMULATION}
\label{sec:detector}
BEPCII is a double-ring $\EE$ collider that has reached a peak luminosity of about $8.5 \times 10^{32}~\rm{cm}^{-2}\rm{s}^{-1}$
at a CM energy of 3.773 GeV.
The cylindrical core of the BESIII detector consists of a helium-based main drift
chamber (MDC), a plastic scintillator time-of-flight (TOF) system, and
a CsI(Tl) electromagnetic calorimeter (EMC), which are all enclosed in
a superconducting solenoidal magnet with a field strength of 1.0~T.
The solenoid is supported by an octagonal flux-return yoke with
resistive plate counter modules interleaved with
steel as muon identifier. The acceptance for charged particles and photons is 93\% over
$4\pi$ stereo angle, and the charged-particle momentum resolution at 1 GeV/$c$ is 0.5\%,
the photon energy resolution at 1.0 GeV is 2.5\% (5\%) in the barrel (end caps).
More details about the apparatus can be found in Ref.~\cite{BESIII}.\\

The response of the BESIII detector is modeled with Monte Carlo (MC) simulations
using a framework based on \textsc{geant}{\footnotesize 4}~\cite{geant4, geant42}.
The production of $\psi$ resonances is simulated with the \textsc{kkmc} generator~\cite{kkmc},
while the subsequent decays are processed via
\textsc{evtgen}~\cite{evt2} according to the branching fractions
provided by the Particle Data Group (PDG)~\cite{PDG2012}, and the
remaining unmeasured decay modes are generated with
\textsc{lundcharm}~\cite{lund}. To determine the detection efficiencies for $\psi\ar\XXB$, $\SSPM$,
 one million MC events are generated for each mode, corresponding to
samples about $20\sim50$ times larger than expected in data.  The events are generated for each channel with our measured angular distribution parameter, which we will introduce in detail later; the $\Xi$ and $\Sigma(1385)$ decays in the signal modes are simulated inclusively according to the corresponding branching fractions taken from PDG~\cite{PDG2012}.

\section{Event selection}
\label{sec:evt_sel}
The selection of $\psi\ar\XXB$, $\SSPM$ events via a full reconstruction of
both $\Xi^{-}(\Sigma(1385)^{\mp})$ and $\bar\Xi^{+}(\bar\Sigma(1385)^{\pm})$ baryons suffers from low
reconstruction efficiency.
To achieve a higher efficiency,
a single baryon $\Xi^{-}$ ($\Sigma(1385)^{\mp}$) tag technique, which does not include the antibaryon mode tag,
is employed to select the signal events $\psi\ar\XXB(\SSPM)$, where only the $\Xi^{-} (\Sigma(1385)^{\mp})$ is reconstructed in
its decay to $\pi^{\mp}\Lambda$ with the subsequent decay $\Lambda \to p\pi^{-}$. Thus,
we require that the events contain at least one positively charged and two negatively charged tracks for the $\XXB (\SSSM)$ channel and two positively charged and one negatively charged track for the $\SSSP$ channel.
Only tracks that are reconstructed in the MDC with good helix fits and within the angular coverage of the
MDC ($|\cos\theta|<0.93$, where $\theta$ is the polar angle with respect to the $e^{+}$ beam direction) are considered.
Information from the specific energy loss measured in MDC ($dE/dx$) and from TOF are combined to form particle identification (PID)
confidence levels for the hypotheses of a pion, kaon, and proton, respectively.
Each track is assigned to the particle type that corresponds to
the hypothesis with the highest confidence level. Events with at least two charged pions ($\pi^{-}\pi^{\mp}$) and at least one proton ($p$) are kept for
further analysis.

In order to reconstruct $\Lambda$ baryons,
a vertex fit is applied to all $p\pi^{-}$ combinations; the ones characterized by $\chi^{2} < 500$ are selected. The invariant mass of the $p\pi^{-}$ pair
is required to be within 6~MeV/$c^{2}$ of the nominal $\Lambda$ mass.
Subsequently, candidates for $\Xi^{-}$ and $\Sigma(1385)^{\mp}$ baryons are built by combining all reconstructed $\Lambda$ with another $\pi^{\mp}$.
The combination
with the minimum $|M_{\pi^{\mp}\Lambda}-M_{\Xi^{-}/\Sigma(1385)^{\mp}}|$ is selected, where $M_{\Xi^{-}/\Sigma(1385)^{\mp}}$ is the nominal mass of $\Xi^{-}$ or $\Sigma(1385)^{\mp}$ from PDG~\cite{PDG2012}.

\begin{figure*}[!hbt]
\bcl
\subfigure{\includegraphics[width=0.36\textwidth]{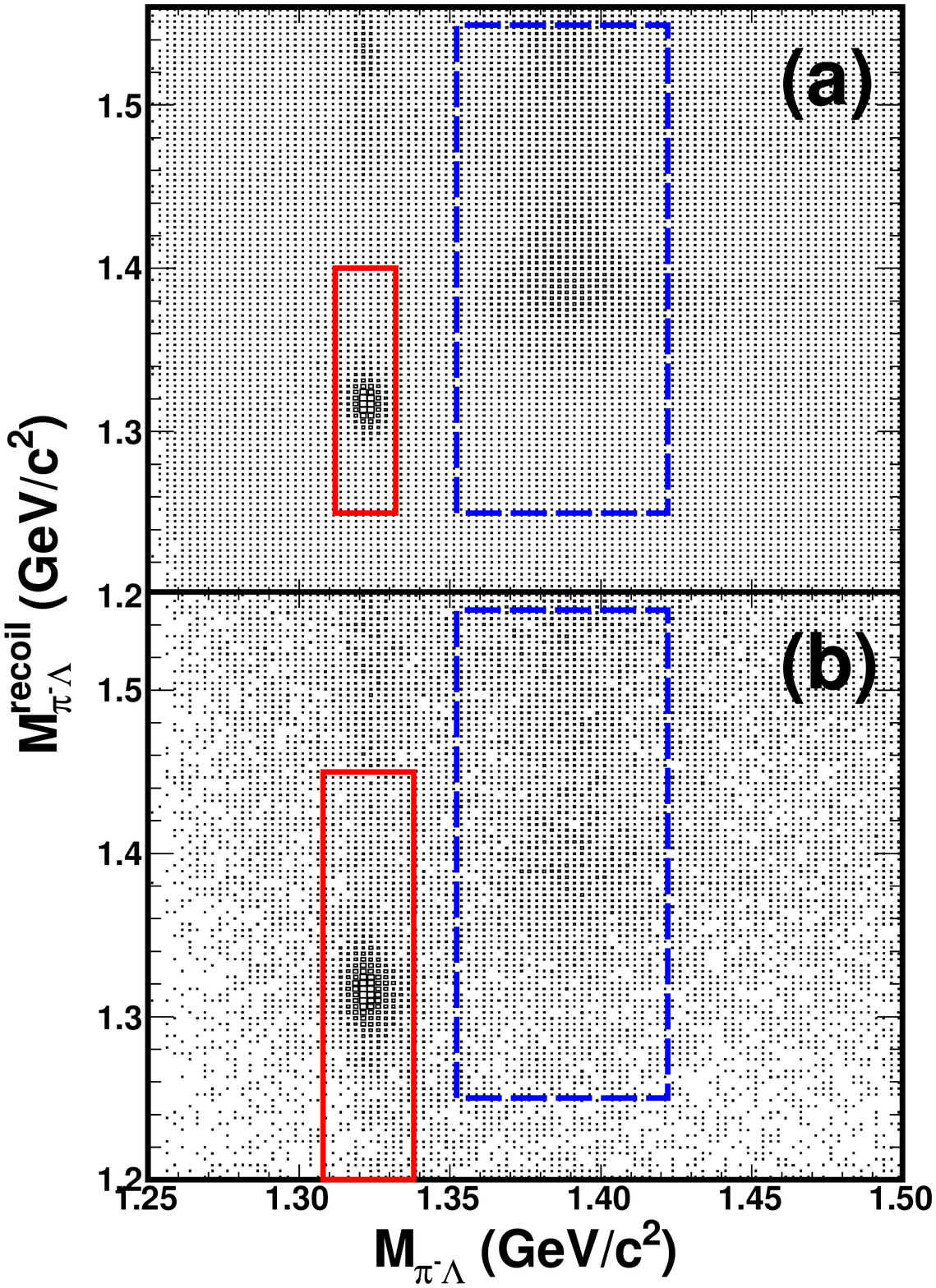}}
\subfigure{\includegraphics[width=0.36\textwidth]{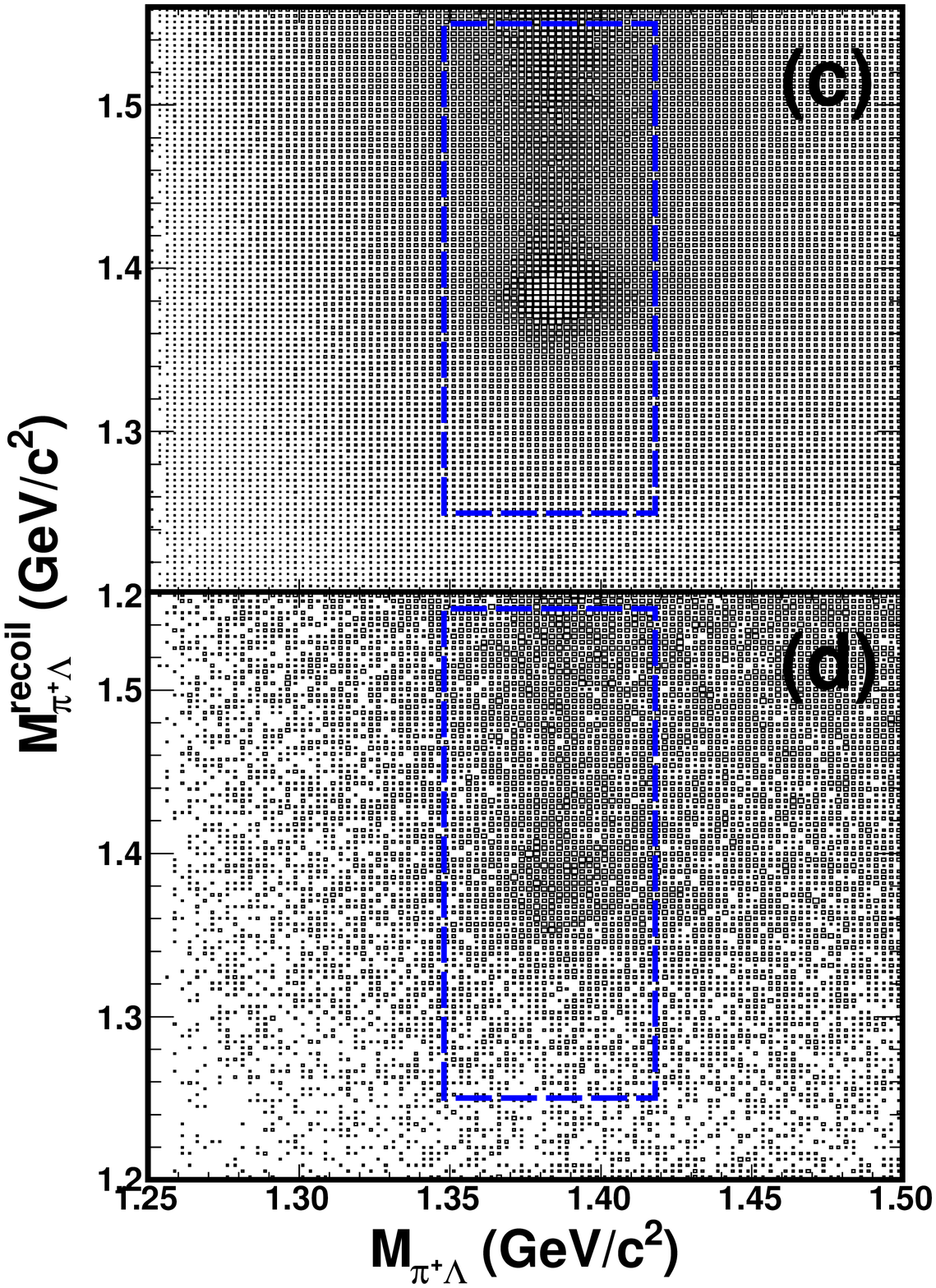}}
\caption{Scatter plots of $M_{\pi^{\pm}\Lambda}$ versus $M^{recoil}_{\pi^{\pm}\Lambda}$ for (a, c) $\jpsi$ and (b, d) $\psp$ data.
The solid boxes are for the $\XXB$ signal region, and the dashed boxes are for the $\SSPM$ signal region.}
\label{scatterplot}
\ecl
\end{figure*}

The partner of $\bar\Xi^{+}$ or $\bar\Sigma(1385)^{\pm}$ is extracted from the mass recoiling against the selected $\pi^{\mp}\Lambda$ system,
\begin{equation}
M^\text{recoil}_{\pi^{\mp}\Lambda} = \sqrt{(E_{CM}-E_{\pi^{\mp}\Lambda})^{2}-\vec{p}^{2}_{\pi^{\mp}\Lambda}},
\end{equation}
where $E_{\pi^{\mp}\Lambda}$ and $\vec{p}_{\pi^{\mp}\Lambda}$ are the energy and the momentum of the
selected $\pi^{\mp}\Lambda$ system, respectively,  and $E_{CM}$ is the $e^+e^-$ CM energy.
Figure~\ref{scatterplot} shows the scatter plots of $M_{\pi^{\mp}\Lambda}$ versus $M^\text{recoil}_{\pi^{\mp}\Lambda}$ for the $\jpsi$ and $\psp$  data samples.
Clear accumulations of events are found for the signals of $\psi \ar \XXB$ ($\SSPM$) decays.
To determine the signal yields, the mass of $\pi^{\mp}\Lambda$ is required to be in the interval
$[1.312, 1.332]$ GeV/$c^{2}$ for $\jpsi\ar\XXB$, and $[1.308, 1.338]$ GeV/$c^{2}$ for $\psp\ar\XXB$, respectively, while we require $|M_{\pi^{\mp}\Lambda}-M_{\Sigma(1385)^{\mp}}|<0.035$ GeV/$c^{2}$ for $\psi\ar\SSPM$.
For the decay $\psp\ar\XXB$ ($\SSSM$), a further requirement of $|M^\text{recoil}_{\pi^{+}\pi^{-}}-M_{\jpsi}| > 0.005$ GeV/$c^{2}$ is applied to
suppress the background $\psp\ar\pi^{+}\pi^{-}\jpsi$, where the $M^\text{recoil}_{\pi^{+}\pi^{-}}$ is the recoil mass of all $\pi^{+}\pi^{-}$ combination, and $M_{\jpsi}$ is the nominal mass of $\jpsi$ according to the PDG~\cite{PDG2012}.

\section{Background study}
\label{sec:background}
Data collected at center-of-mass energies of 3.08 GeV (300 nb$^{-1}$~\cite{Jpsi}) and 3.65 GeV (44 pb$^{-1}$~\cite{Psip}) are used to estimate the
contributions from the continuum processes $\EE\ar\XXB, \SSPM$. After applying the same event selection criteria,
only a few events survive, which do not form any obvious peaking structures
 around the $\bar\Xi^{+}$ or $\bar\Sigma(1835)^{\pm}$ signal regions in  the corresponding $M^\text{recoil}_{\pi^{\mp}\Lambda}$ distribution.
 The scale factor between the data at $\psp$ peak and that at 3.65 GeV is 3.677, taking into account the luminosity and CM energy dependence of the cross section.
This implies that the backgrounds from continuum processes are negligible.

The contamination from other background sources is studied by using MC simulated samples of generic $\psi$ decays that contain the same number of events as data.
After applying the same event selection criteria, it is found that the channels $\jpsi\ar\gamma\eta_{c}$ with $\eta_{c}\ar\XXB$,
$\jpsi\ar\pi^{-}\Lambda\Sigma(1385)^{+}$ (the branching fraction is preliminarily determined with the data based on an iterative method),
and $\jpsi\ar\SSSM$ are potential peaking backgrounds for $\jpsi\ar\XXB$.
According to MC simulations of these backgrounds,
their yields are expected to be negligible after normalization to the total number of $\jpsi$ events. For the $\jpsi\ar\SSPM$ decay, backgrounds are found to be $\jpsi\ar\pi^{\mp}\Lambda\bar\Sigma(1385)^{\pm}$, $\jpsi\ar\Xi(1530)^{-}\bar\Xi^{+} + c.c.$ and $\jpsi\ar\Xi(1530)^{0}\bar\Xi^{0} + c.c.$.
For the $\psp\ar\XXB$ decay, dominant backgrounds come from $\psp\ar\gamma\chicJ$, $\chicJ\ar\XXB$, and $\psp\ar\SSSM$, which are expected to populate smoothly in the $M^\text{recoil}_{\pi^{-}\Lambda}$ spectrum.
For the $\psp\ar\SSPM$ decay, the surviving backgrounds mainly come from the process $\psp\ar\pi^{+}\pi^{-}\jpsi$.

\section{Results}
\subsection{Branching fraction}
\label{sec:branching}
\begin{figure*}[!hbt]
\bcl
\subfigure{\includegraphics[width=0.3\textwidth]{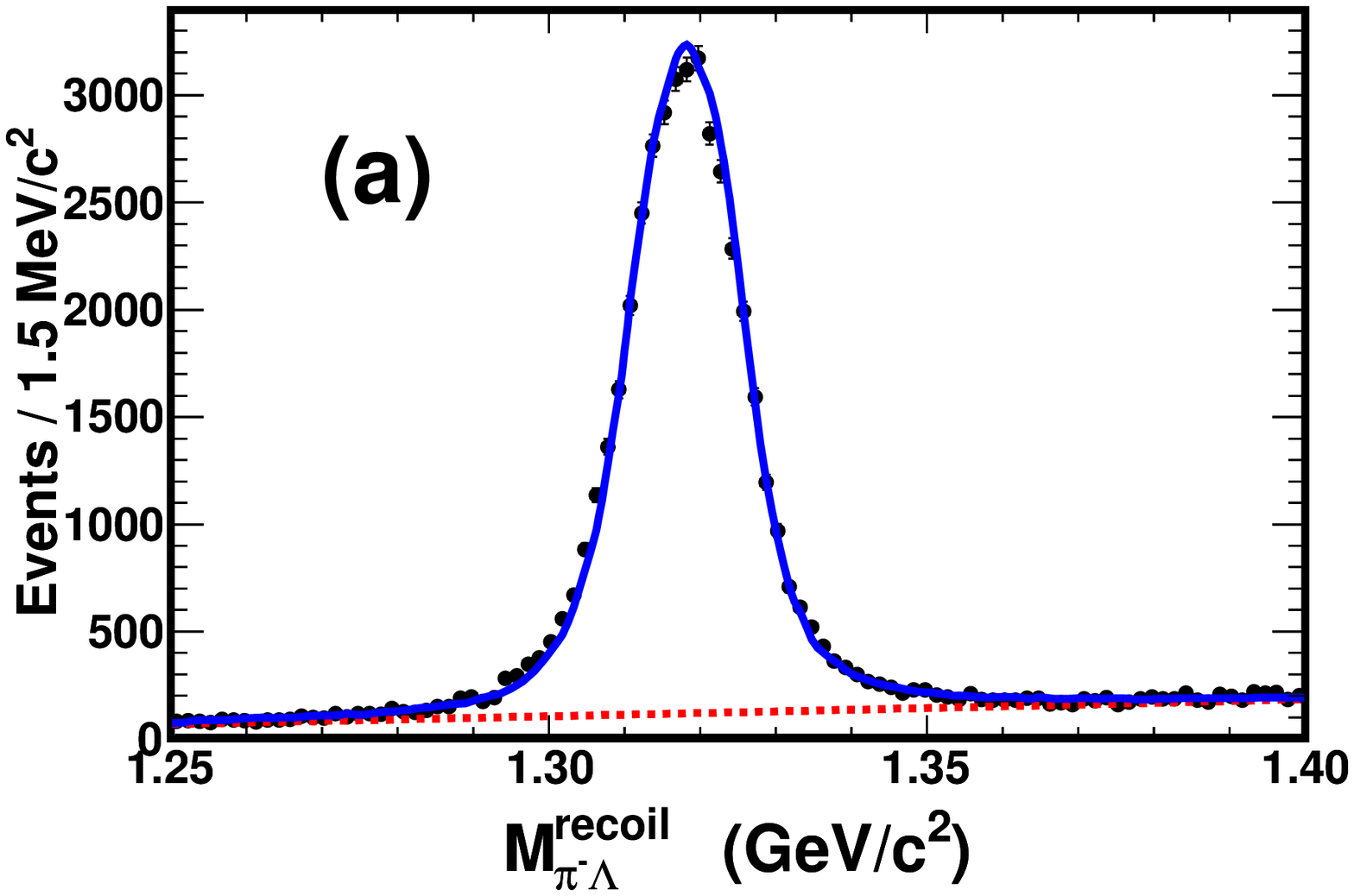}}
\subfigure{\includegraphics[width=0.3\textwidth]{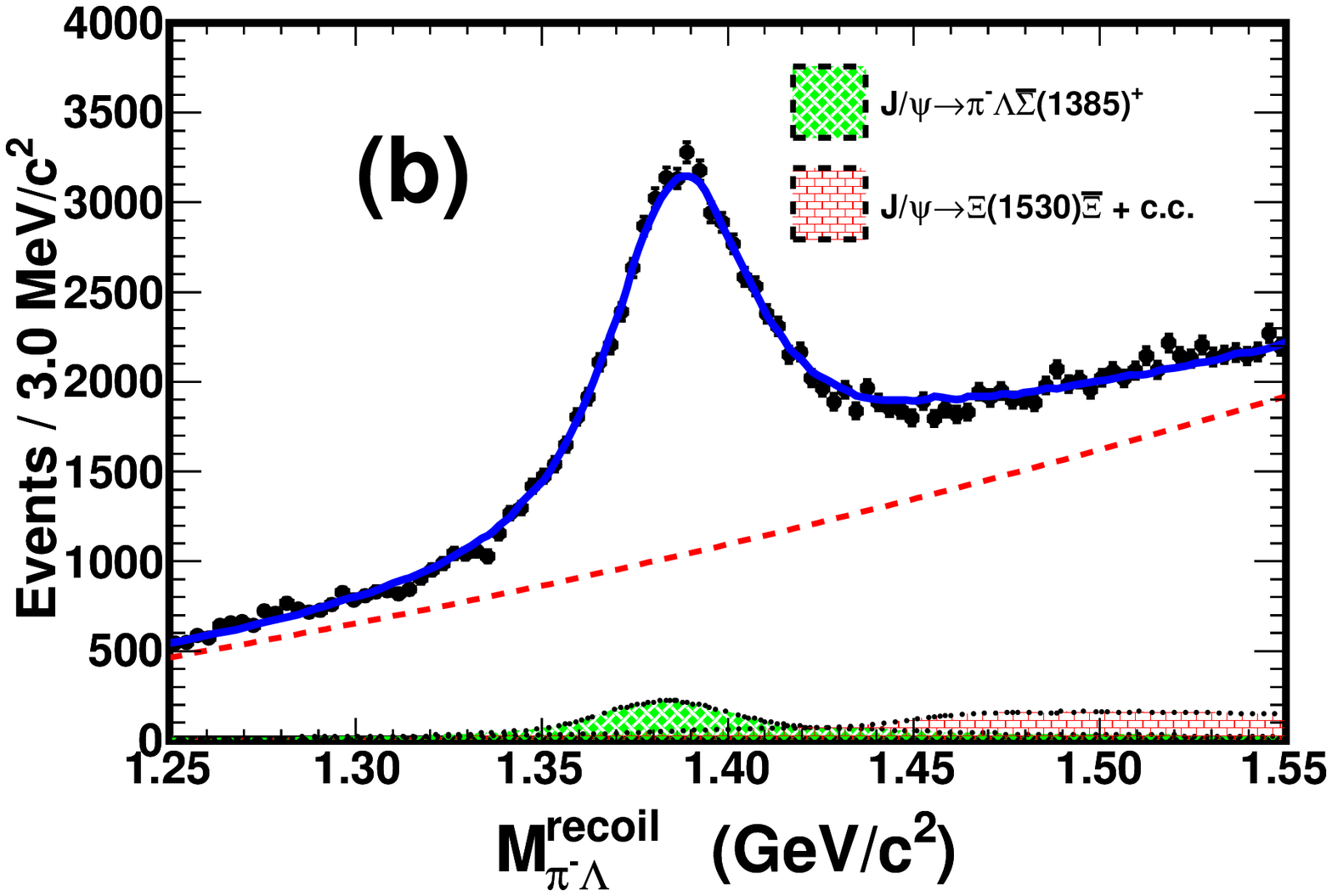}}
\subfigure{\includegraphics[width=0.3\textwidth]{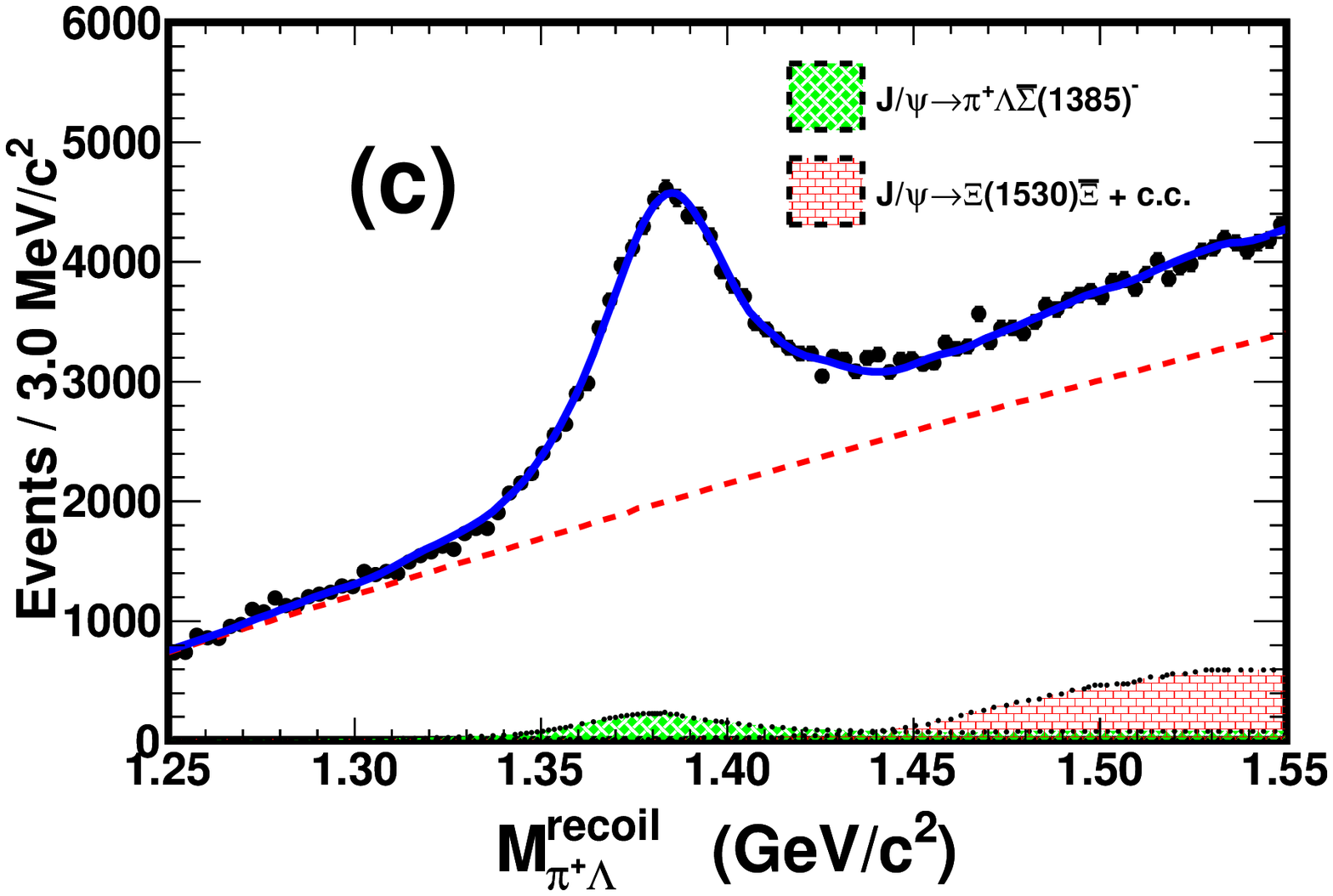}}\\
\subfigure{\includegraphics[width=0.3\textwidth]{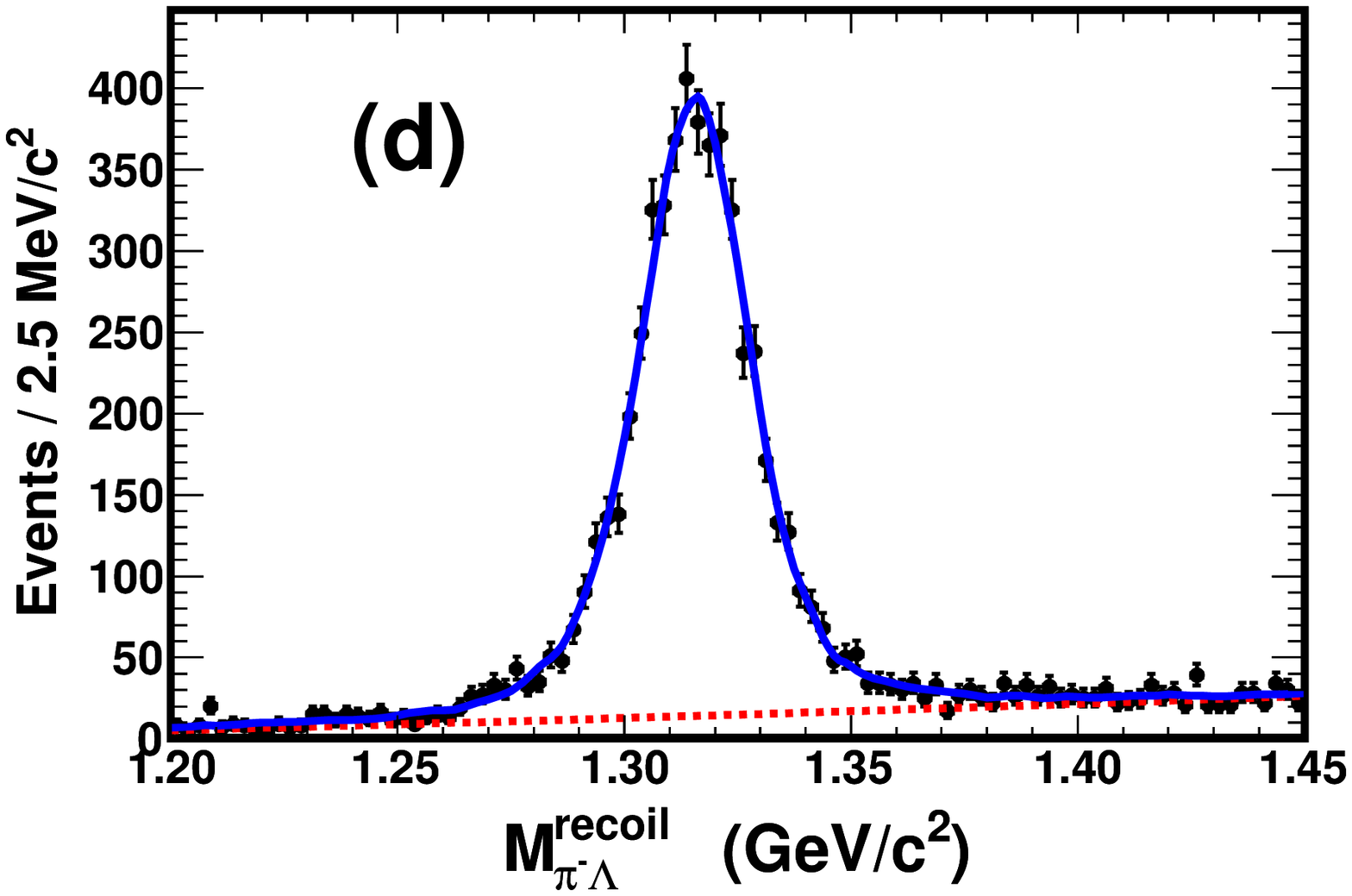}}
\subfigure{\includegraphics[width=0.3\textwidth]{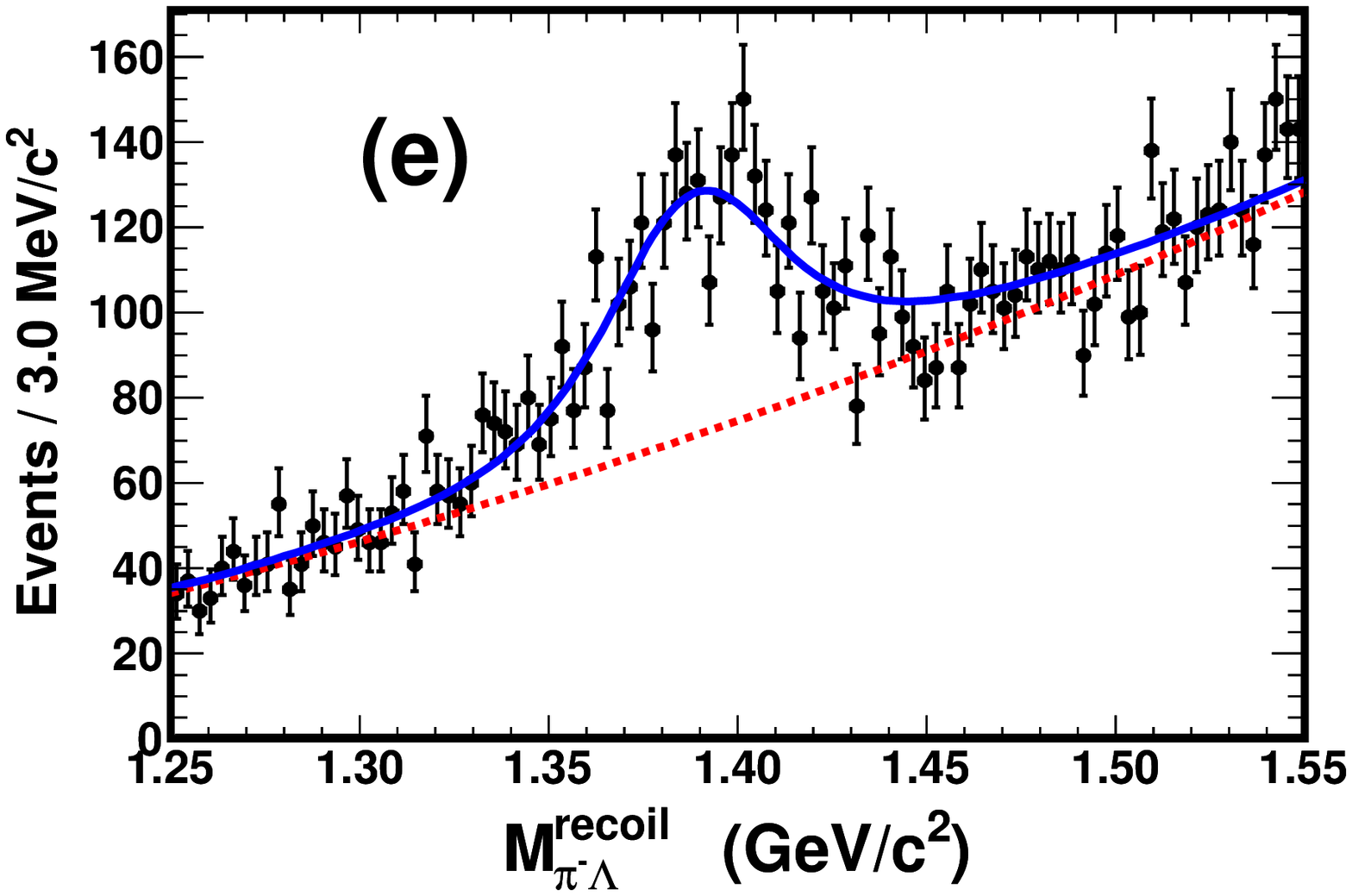}}
\subfigure{\includegraphics[width=0.3\textwidth]{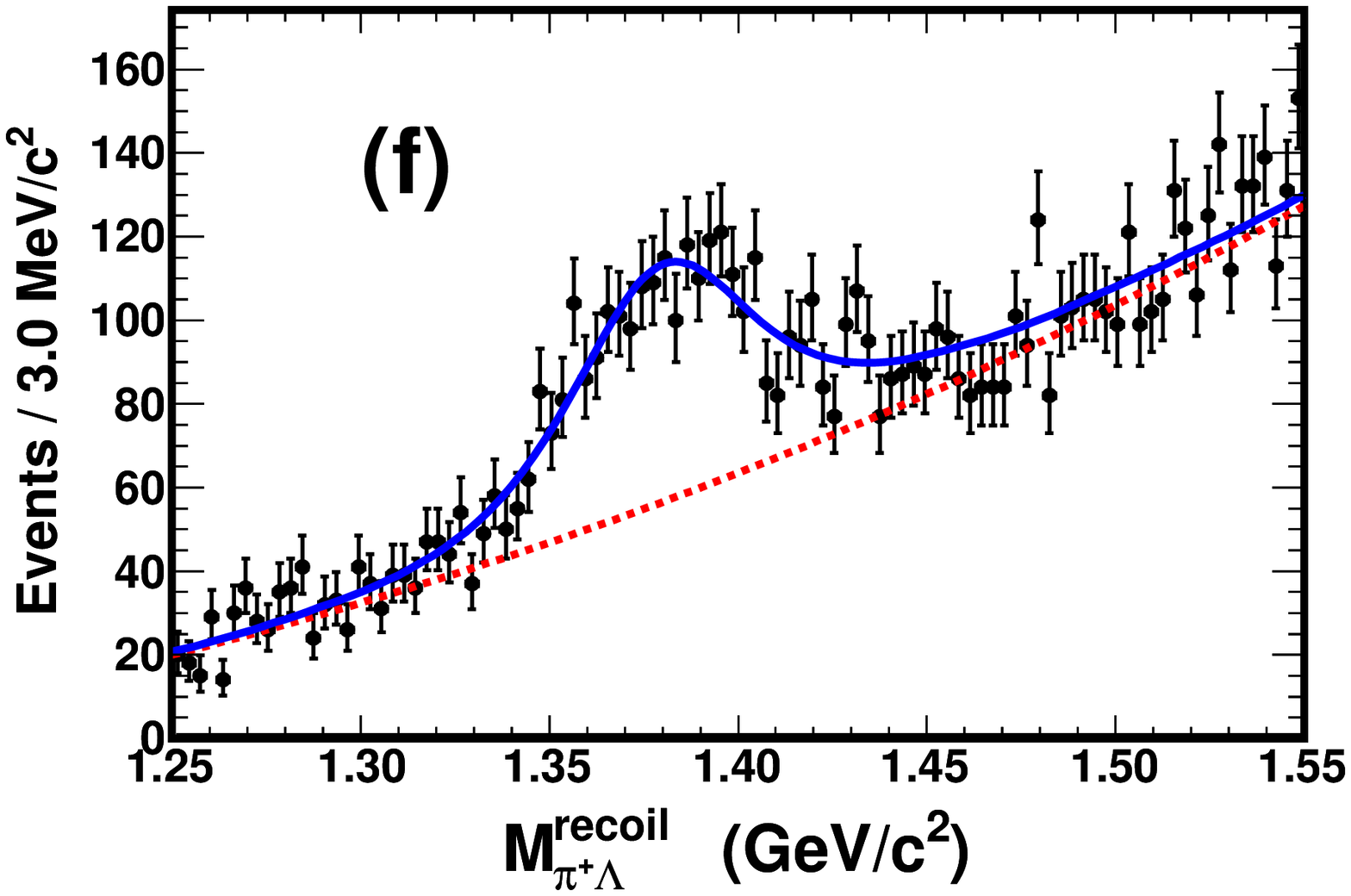}}
%\vspace{-1.0cm}
\caption{Recoil mass spectra of $\pi^{-}\Lambda$ and $\pi^{+}\Lambda$. (a) $\jpsi\ar\XXB$, (b) $\jpsi\ar\SSSM$, (c) $\jpsi\ar\SSSP$, (d) $\psp\ar\XXB$, (e) $\psp\ar\SSSM$ and (f) $\psp\ar\SSSP$. Dots with error bars
indicate the data, the solid lines show the fit results, the dashed lines are for the combinatorial background, and the hatched histograms are for the peaking backgrounds.
}
\label{fitting}
\ecl
\end{figure*}

The signal yields for $\psi\ar\XXB$, $\SSPM$ are determined by performing an extended maximum likelihood fit to $M^\text{recoil}_{\pi^{\mp}\Lambda}$ spectrum. In the fit,
the signal shape is represented by a simulated MC shape convoluted with a Gaussian function taking into account the mass resolution difference between data and MC.
The background shapes for $\psi\ar\XXB$ and $\psp\ar\SSPM$ are represented by a second-order polynomial function since the peaking backgrounds are found to be negligible and the remaining backgrounds are expected to be distributed smoothly in $M^\text{recoil}_{\pi^{\mp}\Lambda}$.
In the decay $\jpsi\ar\SSPM$, the peaking background is found to be significant and is included in the fit. The shapes
of the peaking backgrounds are represented by the individual shapes taken from simulation, and the corresponding number of background events is fixed accordingly. The remaining backgrounds
are described by a second-order polynomial function. Figure~\ref{fitting} shows the projection plots of
$M^\text{recoil}_{\pi^{\mp}\Lambda}$ for $\psi\ar\XXB$ and $\SSPM$.

The branching fractions are calculated by
\begin{equation}
{\cal B}[\psi\ar X]=\frac{N_\text{obs.}}{N_{\psi}\cdot\epsilon},
\end{equation}
where $X$ stands for the $\XXB$ and $\SSPM$ final states, $\epsilon$ denotes the detection efficiencies taking into account the product
branching fraction of the tag mode of $\Xi^{-} (\Sigma(1385)^{\mp})$ decay and the values of $\alpha$ measured in this analysis, $N_\text{obs.}$ is the number of signal events from the fit,
and $N_{\psi}$ is the total number of $\jpsi$ or $\psp$ events~\cite{Jpsi, Psip}. Table~\ref{result} summarizes the number of observed signal events,
the corresponding efficiencies, and branching fractions for the various decays of this measurement with the statistic uncertainty only.
\subsection{Angular distribution}
The values of $\alpha$ for the six decay processes are extracted by performing a least-squares fit to the $\cos\theta$ distributions in the range of 0$.8$ to $-0.8$.  The $\cos\theta$ distributions
are divided into 8 equidistant intervals for the process $\psp\ar\SSPM$ and into 16 intervals for the other four decay modes.

\begin{figure}[!hbt]
\bcl
\subfigure{\includegraphics[width=0.3\textwidth]{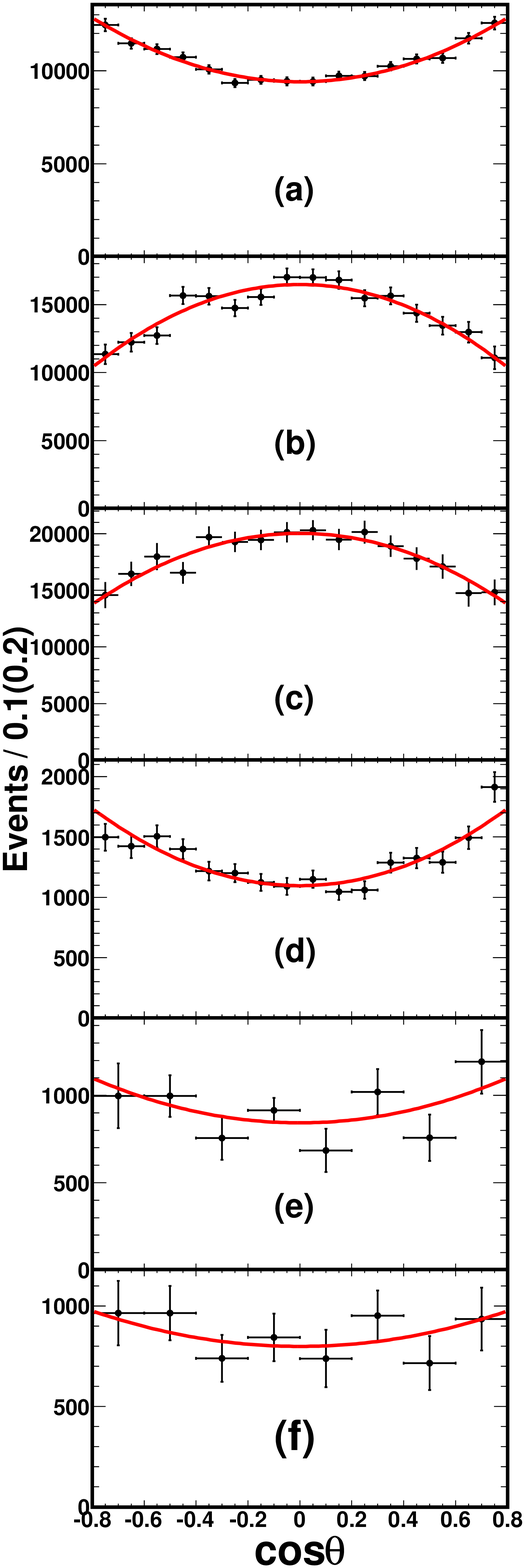}}
\caption{Distributions of $\cos\theta$ for the signals of (a) $\jpsi\ar\XXB$, (b) $\jpsi\ar\SSSM$, (c) $\jpsi\ar\SSSP$, (d) $\psp\ar\XXB$, (e) $\psp\ar\SSSM$ and (f) $\psp\ar\SSSP$. The dots with error bars indicate the efficiency-corrected signal yields in data, and the curves show the fit results.}
\label{angular}
\ecl
\end{figure}

The signal yield in each $\cos\theta$ bin is obtained with the aforementioned fit method.
The distributions of the efficiency-corrected signal yields
together with the curves of the fit are shown in Fig.~\ref{angular}.
The $\alpha$ values obtained from the fits based on Eq.~(\ref{alpha}) are summarized in Table~\ref{result}.

\begin{table*}[!htb]
\bcl
  {\caption{The number of the observed events $N_\text{obs.}$, efficiencies $\epsilon$, $\alpha$ values,
and branching fractions ${\cal B}$ for $\psi\ar\XXB$, $\SSPM$. Only statistical uncertainties are indicated.}\label{result}}
%\scalebox{0.8}[0.9]{
\begin{tabular}{lcccc}  \hline \hline
Channel                                      &$N_\text{obs.}$        &$\epsilon$(\%)     &$\alpha$ &${\cal B} (\times 10^{-4}$)  \\ \hline
$\jpsi\ar\XXB$                               &$42810.7 \pm 231.0 $ & $18.40 \pm 0.04 $  &$~~~0.58 \pm 0.04 $ &$ 10.40 \pm 0.06 $ \\
$\jpsi\ar\SSSM$                              &$42594.8 \pm 466.8 $ & $17.38 \pm 0.04 $  &$-0.58 \pm 0.05$    &$ 10.96 \pm 0.12 $ \\
$\jpsi\ar\SSSP$                              &$52522.5 \pm 595.9 $ & $18.67 \pm 0.04 $  &$-0.49 \pm 0.06$    &$ 12.58 \pm 0.14 $ \\
$\psp\ar\XXB$                                &$5336.7  \pm  82.6 $ & $18.04 \pm 0.04 $  &$~~~0.91 \pm 0.13$  &$~~2.78 \pm 0.05 $ \\
$\psp\ar\SSSM$                               &$1374.5  \pm  97.8 $ & $15.12 \pm 0.04 $  &$~~~0.64 \pm 0.40$  &$~~0.85 \pm 0.06 $ \\
$\psp\ar\SSSP$   &$1469.9 \pm 94.6 $  &$16.45 \pm 0.04 $  &$~~~0.35 \pm 0.37 $  &$~~0.84 \pm 0.05 $ \\ \hline \hline
\end{tabular}%}
\ecl
\end{table*}

\section{Systematic uncertainty}
\label{sec:syst_err}
\subsection{Branching fraction}
Systematic uncertainties on the branching fractions are mainly due to efficiency and resolution differences
between data and MC.  They are
estimated by comparing the efficiencies of tracking, PID, $\Lambda$ and $\Xi^{-}$ reconstruction,
and the $\pi^{\mp}\Lambda$ mass window requirement of the reconstructed
$\Xi(\Sigma(1385)^{\mp})$ between the data and simulation. Additional sources of systematic uncertainties are the
fit range, the background shape, the angular distributions, and the mass shift in $M^{recoil}_{\pi^{\mp}\Lambda}$. In addition, the uncertainties of the decay branching fractions of
intermediate states and uncertainties of the total number of $\psi$ events are also accounted for in the systematic uncertainty.
All of the systematic uncertainties are discussed in detail below.
\begin{enumerate}
    \item The uncertainties due to the tracking and PID efficiencies of the $\pi$ originating from $\Sigma(1385)$ decays are investigated with the
control sample $\jpsi\ar p\bar{p}\pi^{+}\pi^{-}$. It is found that the efficiency difference between data and MC is 1.0\% per pion for
track reconstruction and PID, respectively, taking into account the relative low momentum. These differences are taken as systematic uncertainties.
    \item The uncertainty of the $\Lambda$ reconstruction efficiency in $\Sigma(1385)$ decays is estimated using the control sample $\psi\ar\XXB$. A detailed description of this method can be found
in~\cite{wangxf}. The differences of $\Lambda$ reconstruction efficiency between data and MC are found to be 3.0\% and 1.0\% in the $\jpsi$ and $\psp$ decay respectively, which are taken into account as systematic uncertainties.
    \item The $\Xi$ reconstruction efficiency, which includes the tracking and PID efficiencies for the pion
from the $\Xi$ decay and the $\Lambda$ reconstruction efficiency, is studied with the control
samples $\psi\ar\XXB$ reconstructed via single and double tag methods. The
 selection criteria of the charged tracks,
and the reconstruction of $\Lambda$ and $\Xi$ candidates are exactly the same as those described in Sec.~\ref{sec:evt_sel}.
The $\Xi^{-}$ reconstruction efficiency is defined as the ratio of the number of events from the double tag $\Xi^{-}\bar\Xi^{+}$
to that from the single tag.
        The difference in the $\Xi$ reconstruction efficiency between data and MC samples is taken as the systematic uncertainty.
    \item For $\psi\ar\SSSM$, a strict requirement for the mass window of $\pi^{\mp}\Lambda$ with 1 $\sigma$ level is applied to suppress backgrounds, where the width $\sigma$ of the charged $\Sigma(1385)$ mass is $35\sim40$ MeV~\cite{PDG2012}.
We vary the nominal requirements by $\pm$ 10 MeV/$c^{2}$ and take
the difference between the data and the MC as the systematic uncertainty
due to mass window of $\pi^{\mp}\Lambda$. For the $\Xi$ channels, the systematic uncertainty due to mass window of $\pi^{\mp}\Lambda$ is estimated to be negligible.
    \item In the fits of the $M^\text{recoil}_{\pi^{\mp}\Lambda}$ spectrum, the uncertainty due to the fit range is estimated
by changing the fit range by $\pm$ 10 MeV/$c^{2}$. The differences of the signal yields are taken
as the systematic uncertainties.
    \item The uncertainty related to the shape of nonpeaking backgrounds, which is described by a second-order polynomial function
        in the fit, is estimated by repeating the fit with a first or a third-order polynomial. The largest difference in the signal yield with respect to the nominal yields is taken as the systematic uncertainty.
In the decay $\jpsi\ar\SSPM$, the uncertainty related to the peaking background is estimated
by varying the normalized number of background events by $1 \sigma$. The signal yield changes
are taken as the systematic uncertainty related to the peaking background.
The total uncertainty related to the background are obtained by adding the
individual contributions in quadrature.
    \item The uncertainty in the detection efficiency due to the modeling of the angular distribution of the baryon pairs,
represented by the parameter $\alpha$, is estimated by varying the measured $\alpha$ values by $1\sigma$. The relative change in the detection efficiency is taken as a systematic uncertainty.

\item Due to the imperfection of the simulation of the momentum spectrum of the pion from $\Xi$ or $\Sigma(1385)$ decays, a mass shift ($\sim$2 MeV/$c^2$) between data and MC is observed in the $M^\text{recoil}_{\pi^{\mp}\Lambda}$ spectrum
for the $\jpsi$ decays (the mass shift in $\psp$ decay is negligible), which may affect the signal yields since they are obtained by fitting with the corresponding MC shape convoluted with a Gaussian function.
To estimate the corresponding effect, the shift of the $M^\text{recoil}_{\pi^{\mp}\Lambda}$ spectrum for the simulated exclusive MC events is corrected, and then the data are refitted with the same method as the nominal fit. The resulting changes in signal yields
are taken as the systematic uncertainty.
    \item The uncertainties in the branching fractions of the decays of the intermediate states, $\Xi$, $\Sigma(1385)$ and $\Lambda$, are
taken from PDG~\cite{PDG2012} (0.8\% for $\psi\ar\XXB$ and 1.9\% for $\psi\ar\SSPM$); they are considered as systematic
uncertainties.\\
    \item The systematic uncertainties due to the total number of $\jpsi$ or $\psp$ events are determined with inclusive hadronic $\psi$ decays; they are 0.6\% and 0.8\% for $\jpsi$ and $\psp$~\cite{Jpsi,Psip}, respectively.
\end{enumerate}

The various contributions of the systematic uncertainties on the branching fraction measurements are summarized in Table~\ref{error}. The
total systematic uncertainty is obtained by summing the individual
contributions in quadrature.
\btbl[!htb]
\caption{Systematic uncertainties on the branching fraction measurements (\%).}
\bcl
\doublerulesep 2pt
%\scalebox{0.7}[0.]{
\begin{tabular}{lcccccc}  \hline \hline
\multicolumn{1}{l}{Source} &\multicolumn{3}{c}{$\jpsi\ar$} &\multicolumn{3}{c}{$\psp\ar$} \\ \cline{2-4} \cline{5-7}
Mode                          &$\XXB$ &$\SSSM$ &$\SSSP$  &$\XXB$ &$\SSSM$ &$\SSSP$\\ \hline
MDC tracking                  &--- &1.0  &1.0 &--- &1.0 &1.0\\
PID                           &--- &1.0  &1.0 &--- &1.0 &1.0\\
$\Lambda$ reconstruction      &--- &3.0  &3.0 &--- &1.0 &1.0\\
$\Xi$ reconstruction          &6.6 &---  &--- &4.4 &--- &--\\
Mass window of $\pi\Lambda$  &negligible  &2.1  &1.1 &negligible  &2.4 &2.4\\
Fit range                     &0.2 &2.3  &1.5 &0.2 &3.5 &1.5\\
Background shape              &1.0 &3.6  &4.2 &1.5 &4.5 &4.0\\
Angular distribution          &1.0 &2.0  &1.5 &1.2 &3.0 &2.6\\
Mass shift in $M^{recoil}_{\pi^{\mp}\Lambda}$ &2.0 &1.0 &0.5  &negligible  &negligible &negligible\\
Branching fraction            &0.8 &1.9  &1.9 &0.8 &1.9 &1.9\\
Total number of $\psi$    &0.6&0.6  &0.6  &0.8  &0.8 &0.8\\\hline
Total                         &7.1  &6.5  &6.2  &4.9 &7.4 &6.2\\ \hline \hline
\end{tabular}%}
\label{error}
\ecl
\etbl

\subsection{Angular distribution}
Various systematic uncertainties are considered in the measurement of $\alpha$ values. These include the uncertainty of the signal yield in the
different $\cos\theta$ intervals, the uncertainty of $\cos\theta$ fit procedure, and the uncertainty related to the detection efficiency correction curve as function of $\cos\theta$ bin.
They are summarized in Table~\ref{error_ang} and are discussed in detail below.
\begin{enumerate}
\item The signal yields in each $\cos\theta$ interval are extracted from the
    fit to the corresponding $M^{recoil}_{\pi^{\mp}\Lambda}$ distribution.
    The sources of the systematic uncertainty of the signal yield include
    the fit range, the background shape, and the mass shift in the $M^\text{recoil}_{\pi^{\mp}\Lambda}$ distribution. To estimate the systematic uncertainty related to the fit range on $M^\text{recoil}_{\pi^{\mp}\Lambda}$, we repeat the fit to the $M^\text{recoil}_{\pi^{\mp}\Lambda}$ by changing the fit range
by $\pm$ 10 MeV/$c^{2}$.
Then, the $\alpha$ values are extracted by the fit with the changed signal yields,
and the resulting differences to the nominal $\alpha$ values are taken
as the systematic uncertainties.
Analogously, the uncertainties related to the background shape and the mass shift in $M^\text{recoil}_{\pi^{\mp}\Lambda}$ distribution are evaluated with the method described above.
\item The systematic uncertainties related to the fit procedure of the $\cos\theta$ distributions are estimated by
re-fitting the $\cos\theta$ distribution with a different binning and fit range. We divide $\cos\theta$ into 8 intervals
for $\psi\ar\XXB$, $\jpsi\ar\SSPM$ and 16 intervals for $\psp\ar\SSPM$. The changes of the $\alpha$ values are taken as systematic
uncertainties. We also repeat the fit by changing the range to $[-0.9, 0.9]$ and $[-0.7, 0.7]$ in $\cos\theta$, with the same
 bin size and different number of bins as the nominal fit. The largest difference in $\alpha$ with respect to the nominal value is taken as
the systematic uncertainty.
\item In the analysis, the $\alpha$ values are obtained by fitting the $\cos\theta$ distribution corrected for the detection efficiency. To estimate the systematic uncertainty related to the imperfection of simulation of detection  efficiency, the ratio of detection efficiencies between data and MC simulation is obtained  based on the control sample $\jpsi\ar\XXB$ with a full event reconstruction. Then, the $\cos\theta$ distribution corrected by the ratio of detection efficiencies is refitted. The resulting differences in $\alpha$ are taken as the systematic uncertainty.
\end{enumerate}
All the systematic uncertainties for the $\alpha$ measurement are summarized in Table~\ref{error_ang}. The total systematic
uncertainty is the quadratic sum of the individual uncertainties, assuming them to be independent.
\btbl[!htb]
\caption{Systematic uncertainties on $\alpha$ value measurements (\%).}
\bcl
\doublerulesep 2pt
%\scalebox{0.7}[0.]{
\begin{tabular}{lcccccc}  \hline \hline
\multicolumn{1}{l}{Source} &\multicolumn{3}{c}{$\jpsi\ar$} &\multicolumn{3}{c}{$\psp\ar$} \\ \cline{2-4} \cline{5-7}
Mode                                        &$\XXB$ &$\SSSM$ &$\SSSP$  &$\XXB$ &$\SSSM$ &$\SSSP$  \\ \hline
$M^{recoil}_{\pi^{\mp}\Lambda}$ fitting range&6.6 &5.2 &7.3 &9.1 &7.8 &6.2\\
Background shape                             &5.7 &5.2 &5.9 &7.7 &28.0&11.0\\
Mass shift in $M^{recoil}_{\pi^{\mp}\Lambda}$&4.5 &5.8 &6.0 &negligible &negligible &negligible\\
$\cos\theta$ interval                        &1.5 &2.0 &4.0 &5.6 &16.0&15.0 \\
$\cos\theta$ fit range                       &5.3 &10.5&8.2 &6.6 &25.0&20.0\\
Efficiency correction                        &6.9 &5.1 &5.5 &5.4 &6.1 &6.7\\ \hline
Total                                        &13.2&15.1&15.4&15.7&42.0&28.8\\ \hline \hline
\end{tabular}%}
\label{error_ang}
\ecl
\etbl
\section{Conclusion and discussion}
\label{sec:conclusion}
Using $(225.3 \pm 2.8) \times 10^{6}$ $\jpsi$ and $(106.4 \pm 0.9) \times 10^{6}$ $\psp$ events collected with the BESIII detector at BEPCII,
the branching fractions and the angular distributions for $\psi\ar\XXB$ and $\SSPM$ are measured.
A comparison of the branching fractions and $\alpha$ values between our measurements and previous experiments is summarized in Tables~\ref{result02} and~\ref{result01}, where the branching fractions for
$\psp\ar\SSPM$ and the angular distributions for $\psp\ar\XXB$ and $\SSPM$
are measured for the first time.
The branching fractions and angular distributions for $\jpsi\ar\XXB$, $\SSPM$ and the branching fraction for $\psp\ar\XXB$ are in good agreement and
much more precise compared to previously published results.
The measured $\alpha$ values are also compared with the predictions in theoretical models~\cite{ppbref02, ppbref01}. As indicated in Table~\ref{result01},
most of our results disagree significantly with the
theoretical predictions, which implies that the naive prediction of QCD suffers from the approximation that higher-order corrections are not taken into account.
 The theoretical models are expected to be improved in order to understand the origin of these discrepancies.

To test the \textquotedblleft 12\% rule,\textquotedblright the branching fraction ratios $\frac{{\cal{B}}(\psp\ar\XXB)}{{\cal{B}}(\jpsi\ar\XXB)}$,
$\frac{{\cal{B}}(\psp\ar\SSSM)}{{\cal{B}}(\jpsi\ar\SSSM)}$ and $\frac{{\cal{B}}(\psp\ar\SSSP)}{{\cal{B}}(\jpsi\ar\SSSP)}$  are calculated
to be $(26.73 \pm 0.50 \pm 2.30)\%$, $(7.76 \pm 0.55 \pm 0.68)\%$ and $(6.68 \pm 0.40 \pm 0.50)\%$, respectively, taking into account common systematic uncertainties. The ratios are not in agreement with 12\%, especially for the $\XXB$ mode.
\btbl[!htb]
\caption{Comparison of the branching fractions for $\psi\ar\XXB$, $\SSPM$ (in units of $10^{-4}$). The first uncertainties are statistical, and the seconds are systematic.}
\bcl
\footnotesize{
\doublerulesep 2pt
%\scalebox{0.8}[0.9]{
\begin{tabular}{lcccccc}  \hline \hline
%Mode         &$\jpsi\ar\XXB$ &$\jpsi\ar\SSSM$  &$\psp\ar\XXB$  &$\psp\ar\SSSM$\\ \hline
\multicolumn{1}{l}{Source} &\multicolumn{3}{c}{$\jpsi\ar$} &\multicolumn{3}{c}{$\psp\ar$} \\ \cline{2-4} \cline{5-7}
Mode                     &$\XXB$ &$\SSSM$ &$\SSSP$  &$\XXB$ &$\SSSM$ &$\SSSP$  \\ \hline
This work    &$10.40 \pm 0.06 \pm 0.74$ &$10.96 \pm 0.12 \pm 0.71$ &$12.58 \pm 0.14 \pm 0.78$  &$2.78 \pm 0.05 \pm 0.14$ &$0.85 \pm 0.06 \pm 0.06$&$0.84 \pm 0.05 \pm 0.05$\\
MarkI~\cite{Jximxip01}    &$14.00 \pm 5.00$    &---    &---&$<$ 2.0 &---&---\\
MarkII~\cite{Jximxip02}    &$11.40 \pm 0.80 \pm 2.00$  &$~~8.60 \pm 1.80 \pm 2.20$ &$10.3 \pm 2.4 \pm 2.5$&---   &---&---\\
DM2~\cite{Jximxip03}       &$~~7.00 \pm 0.60 \pm 1.20$ &$10.00 \pm 0.40 \pm 2.10$ &$11.9 \pm 0.4 \pm 2.5$&---  &---&---\\
BESII~\cite{Jximxip04,ppb} &$~~9.00 \pm 0.30 \pm 1.80$ &$12.30 \pm 0.70 \pm 3.00$ &$15.0 \pm 0.8 \pm 3.8$ &$3.03 \pm 0.40 \pm 0.32$  &---&---\\
CLEO~\cite{Pximxip}        &---&---&---&$2.40 \pm 0.30 \pm 0.20$ &---&---\\
BESI~\cite{bes1}           &---&---&---&0.94 $\pm$ 0.27 $\pm$ 0.15&---&---\\
PDG~\cite{PDG2012}    &$8.50 \pm 1.60$         &$10.30 \pm 1.30$ &$10.30 \pm 1.30$&$1.80 \pm 0.60$    &---&---\\  \hline \hline
\end{tabular}}%}
\label{result02}
\ecl
\etbl

\btbl[!htb]
\caption{Comparison of $\alpha$ for $\psi\ar\XXB$ and $\SSPM$.
The first uncertainties are statistical, and the second are systematic.}
\bcl
\footnotesize{
\doublerulesep 2pt
\begin{tabular}{lcccccc}  \hline \hline
\multicolumn{1}{l}{Source} &\multicolumn{3}{c}{$\jpsi\ar$} &\multicolumn{3}{c}{$\psp\ar$} \\ \cline{2-4}  \cline{5-7}
Mode                     &$\XXB$ &$\SSSM$ &$\SSSP$  &$\XXB$ &$\SSSM$ &$\SSSP$  \\ \hline
This work                       &$0.58 \pm 0.04 \pm 0.08$  &$-0.58 \pm 0.05 \pm 0.09$ &$-0.49 \pm 0.06 \pm 0.08$ &$0.91 \pm 0.13 \pm 0.14$ &$0.64 \pm 0.40 \pm 0.27$ &$0.35 \pm 0.37 \pm 0.10$\\
BESII~\cite{Jximxip04}          &$0.35 \pm 0.29 \pm 0.06$  &$-0.54 \pm 0.22 \pm 0.10$&$-0.35 \pm 0.25 \pm 0.06$ &---     &---   &---                          \\
MarkIII~\cite{Jximxip02}        &$0.13 \pm 0.55$     &---  &---&---                      &---       &---       \\        
Claudson  &0.16  &0.11 &0.11 &0.32   &0.29    &0.29      \\
\emph{et al.}~\cite{ppbref02}\\
Carimalo~\cite{ppbref01}&0.27  &0.20  &0.20  &0.52 &0.50     &0.50    \\ \hline \hline
\end{tabular}}
\label{result01}
\ecl
\etbl

\section{acknowledgements}
\label{sec:acknowledgement}
The BESIII Collaboration thanks the staff of BEPCII and the IHEP computing center for their strong support. This work is supported in part by National Key Basic Research Program of China under Contract No. 2015CB856700; National Natural Science Foundation of China (NSFC) under Contracts No. 11125525, No. 11235011, No. 11305180, No. 11322544, No. 11335008, No. 11375205, No. 11425524, No. 11475207, No. 11505034; the Chinese Academy of Sciences (CAS) Large-Scale Scientific Facility Program; the CAS Center for Excellence in Particle Physics (CCEPP); the Collaborative Innovation Center for Particles and Interactions (CICPI); Joint Large-Scale Scientific Facility Funds of the NSFC and CAS under Contracts No. 11179007, No. U1232107, No. U1232201, No. U1332201; CAS under Contracts No. KJCX2-YW-N29, No. KJCX2-YW-N45; 100 Talents Program of CAS; National 1000 Talents Program of China; INPAC and Shanghai Key Laboratory for Particle Physics and Cosmology; German Research Foundation DFG under Contract No. Collaborative Research Center CRC-1044; Istituto Nazionale di Fisica Nucleare, Italy; Koninklijke Nederlandse Akademie van Wetenschappen (KNAW) under Contract No. 530-4CDP03; Ministry of Development of Turkey under Contract No. DPT2006K-120470; Russian Foundation for Basic Research under Contract No. 14-07-91152; The Swedish Resarch Council; U. S. Department of Energy under Contracts No. DE-FG02-05ER41374, No. DE-SC-0010504, No. DE-SC0012069, No. DESC0010118; U.S. National Science Foundation; University of Groningen (RuG) and the Helmholtzzentrum fuer Schwerionenforschung GmbH (GSI), Darmstadt; WCU Program of National Research Foundation of Korea under Contract No. R32-2008-000-10155-0.

\end{document}